\documentclass[12pt]{article}
 \usepackage{epsfig}
  \usepackage{wrapfig}
   \usepackage{graphicx}
   \usepackage{bm}
\begin{document}

\begin{titlepage}
\begin{center}

{\Large\bf Determination of the  fragmentation functions \\[2mm]
from an NLO QCD analysis of the HERMES data \\[2mm]
on pion multiplicities}

\end{center}
\vskip 2cm
\begin{center}
{\bf Elliot Leader}\\
{\it Imperial College London\\ Prince Consort Road, London SW7
2BW, England }
\vskip 0.5cm
{\bf Alexander V. Sidorov}\\
{\it Bogoliubov Theoretical Laboratory\\
Joint Institute for Nuclear Research, 141980 Dubna, Russia }
\vskip 0.5cm
{\bf Dimiter B. Stamenov \\
{\it Institute for Nuclear Research and Nuclear Energy\\
Bulgarian Academy of Sciences\\
Blvd. Tsarigradsko Chaussee 72, Sofia 1784, Bulgaria }}
\end{center}

\vskip 0.3cm
\begin{abstract}
\hskip -5mm

  An NLO QCD analysis of the final HERMES data on pion multiplicities is
  presented and a new set of pion fragmentation functions is extracted
  from the best fit to the data.  We have studied the so-called $[x,z]$
  and $[Q^2,z]$ presentations of their data, as given by HERMES, which,
  in principle, should simply be two different ways of presenting
  the experimental data. We have based our extraction on an excellent
  fit to the $[Q^2,z]$ presentation of the data. We also draw attention 
  to what appears to be a problem with the $[x,z]$ presentation of 
  the HERMES data.


\end{abstract}

\end{titlepage}

\newpage
\setcounter{page}{1}

\section{Introduction}

In the absence of charged current neutrino data, the experiments
on polarized inclusive deep inelastic lepton-nucleon scattering
(DIS) yield information only on the sum of quark and antiquark
parton densities (PDFs), $\Delta q + \Delta \bar{q}$, and the
polarized gluon density $\Delta G$. In order to extract separately
$\Delta q$ and $\Delta \bar{q}$ other reactions are needed. One
possibility is to use the {\it polarized} semi-inclusive
lepton-nucleon processes (SIDIS) $l+ N \rightarrow l'+h+X$, where
$h$ is a detected hadron (pion, kaon, etc) in the final state. In
these processes new physical quantities appear - the collinear
fragmentation functions $D^h_{q, \bar q}(z, Q^2)$ which describe
the fragmentation of quarks and antiquarks into hadrons. Due to
the different fragmentation of quarks and antiquarks, the
polarized parton densities $\Delta q$ and $\Delta \bar{q}$ can be
determined separately from a combined QCD analysis of the data on
inclusive and semi-inclusive asymmetries. The key role of the
fragmentation functions for the correct determination of sea quark
parton densities $\Delta \bar{q}$, especially of the polarized
strange quark densitiy, was discussed in \cite{deltas_puzzle}.
Note that the W data from RHIC give no information about the
polarized strange quark density and cannot help to solve the so
called "strange quark polarization puzzle" (see the second
reference in [1]).

There are different sources to extract the fragmentation functions
(FFs) themselves: semi-inclusive $e^+ \, e^-$ annihilation data,
single-inclusive production of a hadron $h$ at a high transverse
momentum $p_T$ in hadron-hadron collisions, unpolarized
semi-inclusive DIS processes. It is important to mention that the
data on hadron multiplicities in unpolarized SIDIS processes are
crucial for a reliable determination of FFs, because only then can
one separate $D_q^h(z,Q^2)$ from $D_{\bar q}^h(z,Q^2)$ (from the
other processes only the sum of them can be determined). The first
global analysis based on all these reactions was carried out by de
Florian, Sassot, Stratmann (DSS) group \cite{DSS}. As a result,
the properties of the extracted set of FFs significantly differed,
especially in the kaon sector, from those of the other then
published sets of FFs \cite{other_FFs} determined from analyses in
which the SIDIS data have been not included. Unfortunately, the
DSS FFs were based on the {\it unpublished} HERMES'05 SIDIS data
on hadron multiplicities \cite{HERMES2005}, which were not
confirmed in the final HERMES data \cite{HERMES}. Indeed the final
HERMES data differ significantly from those used in the analysis
of \cite{DSS} so that the FFs extracted in \cite{DSS} are
incorrect (see Figures 9 and 10 in \cite{HERMES} for LO DSS FFs,
and fig. 5 in \cite{newDSS_FFs} for NLO DSS FFs). Moreover, in the
extraction of the next-to-leading order (NLO) DSS set of FFs there
was a mistake (see the correction in the Appendix in
\cite{Vogelsang}) in the expression for the longitudinal gluon
Wilson coefficient function in the theoretical formulae for the
multiplicities (we, independently, became aware of this error
recently). It has turned out that not only the DSS FFs, but all
the other sets of pion and kaon FFs presented in \cite{other_FFs}
are NOT in agreement with the final HERMES \cite{HERMES} and the
preliminary COMPASS data \cite{COMPASS} on hadron multiplicities.
In our paper \cite{DSPIN'13} a theoretical analysis of these data
was performed and new  sets of pion fragmentation functions were
extracted from the best NLO QCD fits to the data, and it was shown
that they disagree significantly with the pion FFs determined from
all previous analyses. Very recently de Florian at al. (DSEHS)
have presented results on pion FFs obtained from their \emph{new}
global QCD analysis \cite{newDSS_FFs} using the final HERMES and
the preliminary COMPASS data on pion multiplicities.

In our paper \cite{DSPIN'13} we pointed out a possible
inconsistency between the HERMES $[x, z]$ and $[Q^2, z]$
presentations of their data on pion multiplicities. Bearing in
mind that the semi-inclusive DIS hadron production processes are
essential for the separation of $D_q^{h}$ and $D_{\bar q}^{h}$
fragmentation functions, we present in this paper a more detailed
discussion of our previous analysis in which we have taken into
account the mistake in the longitudinal gluon Wilson coefficient
function \cite{Double_Mellin} present in our previous analysis,
and have used the corrected version given in \cite{Vogelsang}.
Also, instead of the NLO MRST'02 set \cite{MRST02} we have here
utilized the newer NLO MSTW'08 set \cite{MSTW08} of unpolarized
parton densities and study the influence of this on the extracted
FFs.

\section{QCD treatment of pion multiplicities}

The multiplicities $M_{p(d)}^{\pi}(x,Q^2,z)$ of pions using a
proton (deuteron) target are defined as the number of pions
produced, normalized to the number of DIS events, and can be
expressed in terms of the semi-inclusive cross-section
$\sigma_{p(d)}^{\pi}$ and the inclusive cross-section
$\sigma_{p(d)}^{DIS}$:
\begin{eqnarray}
M_{p(d)}^{\pi}(x,Q^2,z)&=&
\frac{d^3N_{p(d)}^{\pi}(x,Q^2,z)/dxdQ^2dz}{d^2N_{p(d)}
^{DIS}(x,Q^2)/dxdQ^2}\Leftrightarrow\frac{d^3\sigma_{p(d)}^{\pi}(x,Q^2,z)
/dxdQ^2dz}{d^2\sigma_{p(d)}^{DIS}(x,Q^2)/dxdQ^2}\nonumber\\
&=&\frac{(1+(1-y)^2)2xF_{1p(d)}^{\pi}(x,Q^2,z)+2(1-y)xF_{Lp(d)}
^{\pi}(x,Q^2,z)}
{(1+(1-y)^2)2xF_{1p(d)}(x,Q^2)+2(1-y)F_{Lp(d)}(x,Q^2)}.
\label{M_exp_th}
\end{eqnarray}
In Eq. (\ref{M_exp_th}) $F_1^{\pi}, F_L^{\pi}$ and $F_1, F_L$ are
the semi-inclusive and the usual nucleon structure functions,
respectively. $F_1^{\pi}$ and $ F_L^{\pi}$ are expressed in terms
of the unpolarized parton densities and fragmentation functions
(see \cite{DSS}), while $F_1$ and $F_L$ are given purely in terms of the
unpolarized parton densities.

We have assumed in our analysis that  isospin $SU(2)$ symmetry
for the favored and unfavored fragmentation functions holds
\begin{equation}
D^{\pi^{+}}_{u}(z, Q^2_0)=D^{\pi^{+}}_{\bar d}(z, Q^2_0),~
D^{\pi^{+}}_{\bar u}(z, Q^2_0)=D^{\pi^{+}}_{d}(z, Q^2_0),
\end{equation}
and in addition, the following relations for the fragmentation of
strange quarks into a pion:
\begin{equation}
D^{\pi^{+}}_{s}(z, Q^2_0)=D^{\pi^{+}}_{\bar s}(z,
Q^2_0)=D^{\pi^{+}}_{\bar u}(z, Q^2_0). \label{s_eq_ub}
\end{equation}
Due to the charge conjugation invariance of the strong
interactions the fragmentation functions $D^{\pi^{-}}_{q, \bar q}$
can be expressed through $D^{\pi^{+}}_{q, \bar q}$:
\begin{equation}
D^{\pi^{-}}_{q(\bar q)}(z, Q^2_0)=D^{\pi^{+}}_{\bar q (q)}(z,
Q^2_0),~~ D^{\pi^{-}}_{g}(z, Q^2_0)=D^{\pi^{+}}_{g}(z, Q^2_0).
\end{equation}

As a result, we have to extract only three independent FFs
($D^{\pi^{+}}_{u},~ D^{\pi^{+}}_{\bar u},~ D^{\pi^{+}}_{g}$) from
the NLO QCD fit to HERMES proton and deuteron data on pion
multiplicities. The charm contribution to the multiplicities is
not taken into account. In the theoretical analysis of the data
the Mellin transform technique \cite{Double_Mellin} was used to
calculate the semi-inclusive $F^h_{1,L}(x,Q^2,z)$ and the usual
$F_{1,L}(x,Q^2)$ nucleon structure functions in
Eq.(\ref{M_exp_th}) from their moments. The expressions for the
moments of the Wilson coefficient functions $C_{ij}^{(1)}(x,z)$
needed in these calculations can been found in
\cite{Double_Mellin}. As was mentioned in the Introduction, the
error in the gluon Wilson coefficient, $C^{(1),nm}_{L,qg}$, was
corrected. Compared to our previous fit \cite{DSPIN'13} where for
the unpolarized PDFs we have used the NLO MRST'02 set
\cite{MRST02}, we use now the NLO MSTW'08 set \cite{MSTW08}, for
which the strange quark density $s(x,Q^2)$ is not equal to $\bar
s(x,Q^2)$. Note that we have chosen this set of PDFs in order to
be able to compare correctly our extracted pion FFs with those of
DSEHS obtained from the recent global fit \cite{newDSS_FFs} where
the MSTW'08 set of PDFs has been used. The influence of the choice
of the unpolarized densities on the extracted FFs will be
discussed.

For the input FFs the following parametrization at $Q^2_0=1~GeV^2$
was used:
\begin{equation}
zD^{\pi^{+}}_{i}(z,
Q^2_0)=\frac{N_iz^{\alpha_i}(1-z)^{\beta_i}[1+\gamma_i
(1-z)^{\delta_i}]}{B[\alpha_i+1,\beta_i+1]+
\gamma_iB[\alpha_i+1,\beta_i+\delta_i+1]}, \label{inputFFs}
\end{equation}
where the parameters $\{N_i,\alpha_i,\beta_i,\gamma_i,\delta_i\}$
are free parameters to be determined from the fit to the data.
Here, $i$ stands for $u, \bar u$ and $g$, while $B(a,b)$ denotes
the Euler beta function, and the $N_i$ are chosen in such a way that
they represent the contribution of $zD^{\pi^{+}}_i$ to the
momentum sum rule.

\section{Results of analysis}

Let us discuss now our results on the pion FFs extracted from our
NLO QCD fit to the HERMES proton and deuteron data on pion
multiplicities, corrected for exclusive vector meson production
\cite{HERMES}. In our study we have analyzed the $[Q^2,z]$ and
$[x,z]$ presentations of the data (see Fig. 8 in \cite{HERMES},
left column, second and third lines) for which the multiplicities
do not depend on $P_{h\perp}$, were $P_{h\perp}$ is the component
of the hadron momentum, $P_h$, transverse to the momentum of the
virtual photon. They correspond to two-dimensional projections
obtained by the HERMES group from the full HERMES data sets
$[Q^2,z, P_{h\perp}]$ and $[x,z,P_{h\perp}]$, respectively. The
pion multiplicities are given for 4 z-bins
[0.2-0.3;~0.3-0.4;~0.4-0.6;~0.6-0.8] as functions of the mean
value of $Q^2$, $<Q^2>$, of each individual $Q^2$ bin for the
$[Q^2,z]$ presentation or as functions of the mean value of $x$,
$<x>$, of each individual $x$ bin for the $[x,z]$ presentation.
Note that for the $[Q^2,z]$ presentation there is no binning in
$x$. This means that the multiplicity measured in a given $Q^2$
bin, $Q^2_{min}\leq Q^2 \leq Q^2_{max}$, corresponds to the
summing over all possible values of $x$ belonging to the strip in
the $\{x-Q^2\}$ plane, bounded by $Q^2_{min},~Q^2_{max}$ and the
kinematics of the HERMES experiment. And vice versa, for the
$[x,z]$ presentation, there is no binning in $Q^2$, and the
multiplicity measured for each $x$ bin corresponds to all possible
values of $Q^2$ belonging to the $\{x,Q^2\}$ strip fixed by the
boundaries of the $x$ bin and the kinematics of the HERMES
experiment. Thus, in principle, in the theoretical calculation of
the pion multiplicities one has to integrate the semi-inclusive
and inclusive cross section on RHS side of Eq. (\ref{M_exp_th})
over the $x$ and $Q^2$ regions corresponding to each $Q^2$ bin for
the $[Q^2,z]$ presentation or to each $x$ bin for the $[x,z]$ one.
It turns out however, that replacing $x$ and $Q^2$ by their mean
values $<x>$ and $<Q^2>$ in the calculation of the multiplicities
leads to very small difference. Further details are presented
later.
\begin{figure} [h]
\begin{center}
\includegraphics[height=.40\textheight]{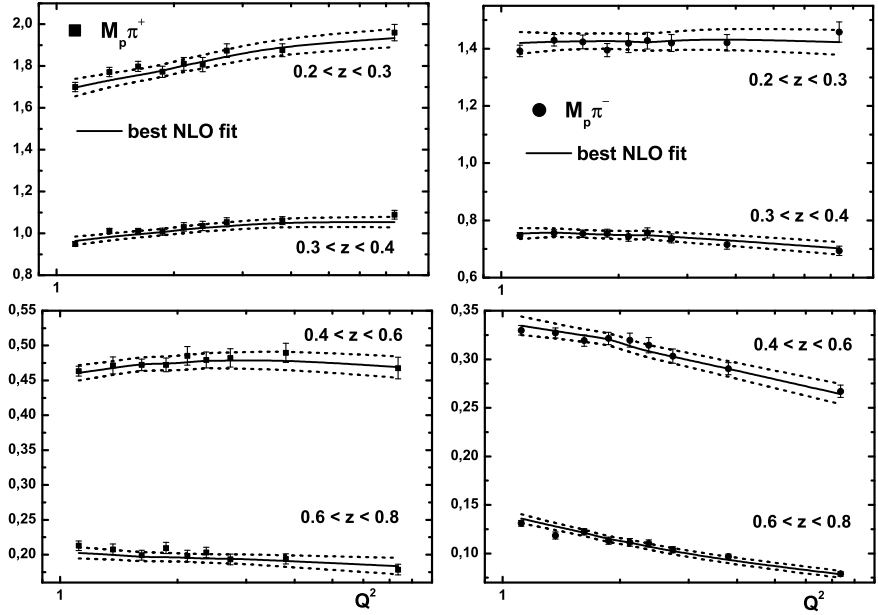}
\caption{\footnotesize Comparison of HERMES $[Q^2,z]$ {\it proton}
data on
$\pi^{+}$ (left) and $\pi^{-}$ multiplicities (right) with the
best NLO fit curves. The error bands (the area between the dot
curves) correspond to uncertainty estimates at 68\% C.L. The
errors of the data are {\it total}, statistical and systematic
taken in quadrature. \label{fig1} }
\end{center}
\end{figure}

\begin{figure}
\begin{center}
  \includegraphics[height=.40\textheight]{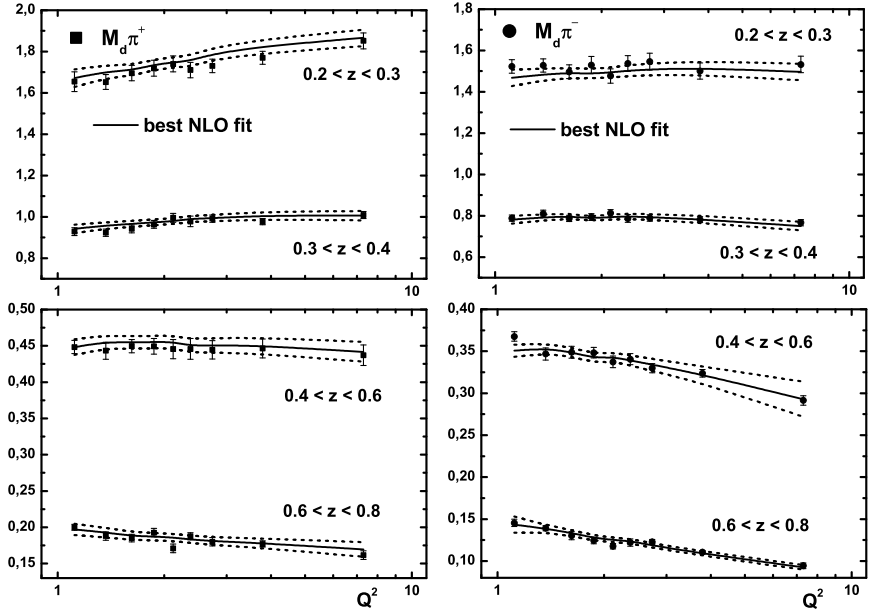}
\caption{\footnotesize Comparison of HERMES $[Q^2,z]$ {\it deuteron} data on
$\pi^{+}$ (left) and $\pi^{-}$ multiplicities (right) with the
best NLO fit curves. The error bands (the area between the dot
curves) correspond to uncertainty estimates at 68\% C.L. The
errors of the data are {\it total}, statistical and systematic
taken in quadrature. \label{fig2} }
\end{center}
\end{figure}

The total number of the $\pi^{+}$ and $\pi^{-}$ data points for
each of the presentations is 144, 72 for $\pi^{+}$ and 72 for
$\pi^{-}$ data. In the case of $[Q^2,z]$ presentation of the data
a good fit to the proton and deuteron data is achieved,
$\chi^2/{\rm d.o.f}$ = 123.95/132 = 0.94 for 144 experimental
points and 12 free parameters. The errors used in the fit are
quadratic combinations of the statistical and point-to-point
systematic errors. We have found that the description of the
proton data (the mean value of $\chi^2$ per point is equal to 0.83
for $\pi^{+}$ and 0.65 for $\pi^{-}$ multiplicities) is better
than that of the deuteron data (where the mean value of $\chi^2$
per point is equal to 0.98 for $\pi^{+}$ as well as for $\pi^{-}$
multiplicities). The quality of the fit to the data is illustrated
in Fig.~\ref{fig1} (for the proton target) and Fig.~\ref{fig2}
(for the deuteron target). The error bands (the area between the
dot curves) correspond to uncertainty estimates at 68\% C.L. Note
that the vertical scale is linear, not logarithmic.
\vskip 0.5cm
\begin{center}
\begin{tabular}{cl}
&{\bf TABLE I.} The parameters of the NLO input FFs at $Q^2=1~GeV^2$ obtained\\
&from the best fit to the data. The parameters marked by
(*) are fixed.
\end{tabular}
\vskip 0.3 cm
\begin{tabular}{|c|c|c|c|c|c|c|} \hline
  Flavor &  N &  $\alpha$ &  $\beta$ & $\gamma$ & $\delta$  \\ \hline
 $u$& 0.278~$\pm$~0.016 & 0.276~$\pm$~0.192 & 0.188~$\pm$~0.187 &
 7.64~$\pm$~1.62~ & 3.09~$\pm$~0.40\\
 $\bar{u}$& 0.153~$\pm$~0.016 & 0.282~$\pm$~0.251 & 1$^*$ &
 9.18~$\pm$~4.12 & 3.85~$\pm$~0.45 \\
g & 0.113~$\pm$~0.005 & 12.70~$\pm$~5.64 & 14.39~$\pm$~6.35 &
0$^*$ & --\\
\hline
\end{tabular}
\end{center}
\vskip 0.5cm

The values for the parameters of the input FFs (\ref{inputFFs})
obtained from the best fit to the data are presented in Table I.
It turned out during the fit that there was a slight preference
for the parameter $\beta_{\bar u}$ to go to the somewhat
unphysical limit zero, but the value of $\chi^2$, as well as the
values of $D^{\pi^{+}}_{\bar u}(z)$ for the measured range of z,
$z\in[0.2,~0.8]$, practically do not change for  fixed values of
$\beta_{\bar u}$ in the range $[0,~ 2]$. That is why it was fixed
at the reasonable value $\beta_{\bar u}=1$. Also, because of the
small $Q^2$ range of the HERMES data, a simpler parametrization
for the gluon FF $D^{\pi^{+}}_g(z)$ was used with only three
parameters and $\gamma_g=0$.

\begin{figure}
\begin{center}
  \includegraphics[height=.46\textheight]{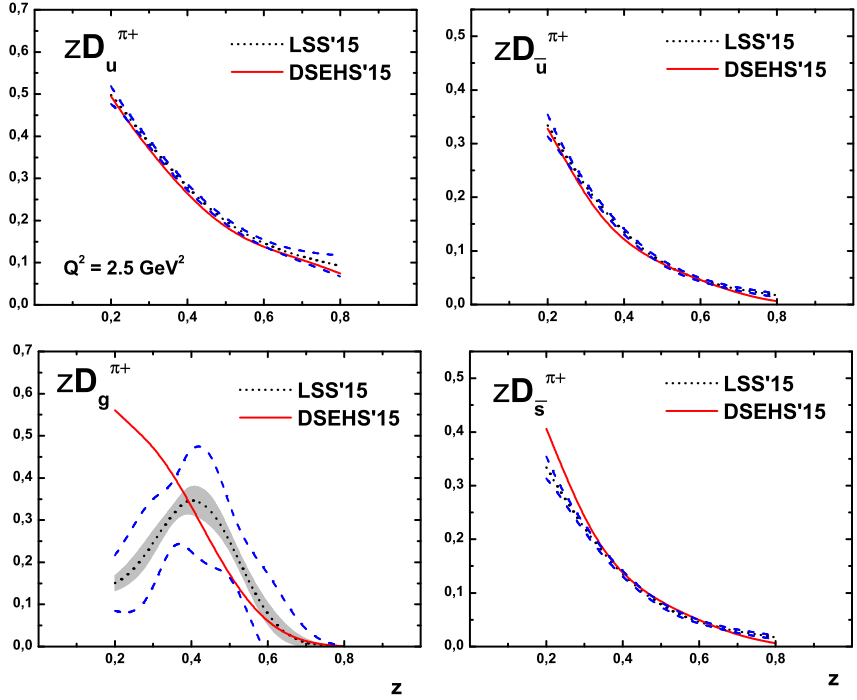}
\caption{\footnotesize Comparison between our pion FFs at $Q^2=2.5~\rm GeV^2$ along with the uncertainty estimates at 68\% C.L. (the area
between the dashed curves) and those of DSEHS (solid curves). For
the gluon fragmentation function the uncertainty corresponding to
$\Delta \chi^2=1$ (the black shaded band) is also presented.
\label{fig3} }
\end{center}
\end{figure}

The extracted pion FFs from the fit to HERMES $[Q^2,z]$ data on
pion multiplicities are presented in Fig. 3 along with their error
bands corresponding to the uncertainty estimates at 68\% C.L, and
compared to those determined recently by DSEHS from their global
analysis \cite{newDSS_FFs} which also made use of the HERMES
$[Q^2,z]$ data. In Fig. 3 the error band for the gluon
fragmentation function corresponding to $\Delta \chi^2=1$ (the
black shaded band) is also presented. The corresponding error
bands for the other FFs are not presented because they are very
narrow and practically not visible. The fragmentation functions
are plotted for the mean value of $Q^2$ for the HERMES data,
$Q^2=2.5~\rm GeV^2$, and for the measured z region [0.2-0.8]. One
can see from Fig.~\ref{fig3} that our (LSS) pion FFs
$D^{\pi^{+}}_{u}(z)$ and $D^{\pi^{+}}_{\bar u}(z)$ are close to
those of DSEHS (solid curves). In the DSEHS analysis the equality
(\ref{s_eq_ub}) for the fragmentation of the strange and $\bar u$
quarks into pion is not assumed. As a result, the extracted
fragmentation functions for the strange quark, $D^{\pi^{+}}_{\bar
s}(z)$, differ a little in the z range $0.2 < z < 0.35$. The main
difference between the extracted FFs is for the gluons. This is
not unexpected bearing in mind that an accurate determination of
the qluon fragmentation function requires data covering a large
range  in $Q^2$ and that for the semi-inclusive DIS processes the
range for the HERMES $[Q^2,z]$ data is small: $1.1 < Q^2 < 7.4~\rm
GeV^2$.

\begin{figure}
\begin{center}
  \includegraphics[height=.26\textheight]{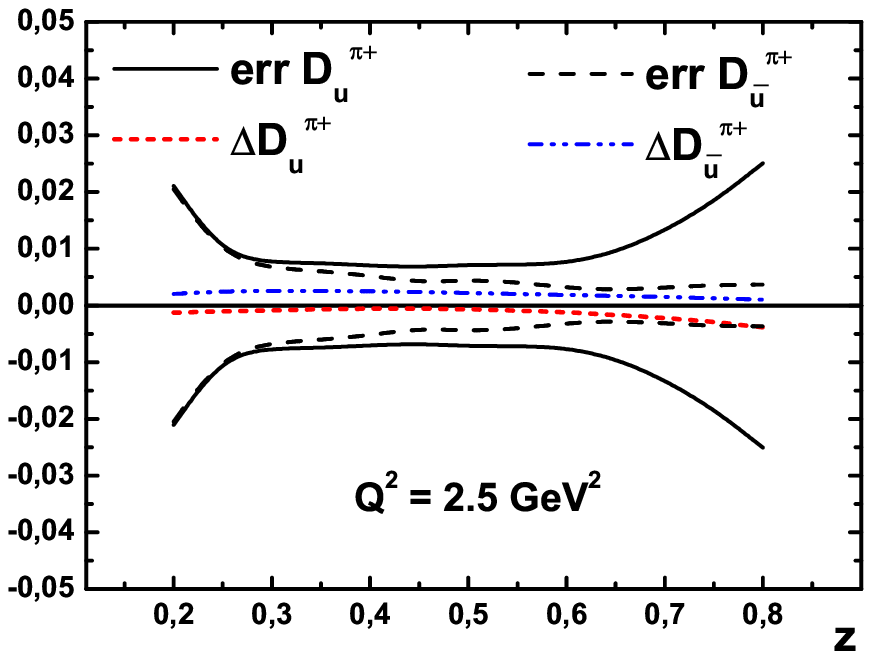}
\includegraphics[height=.26\textheight]{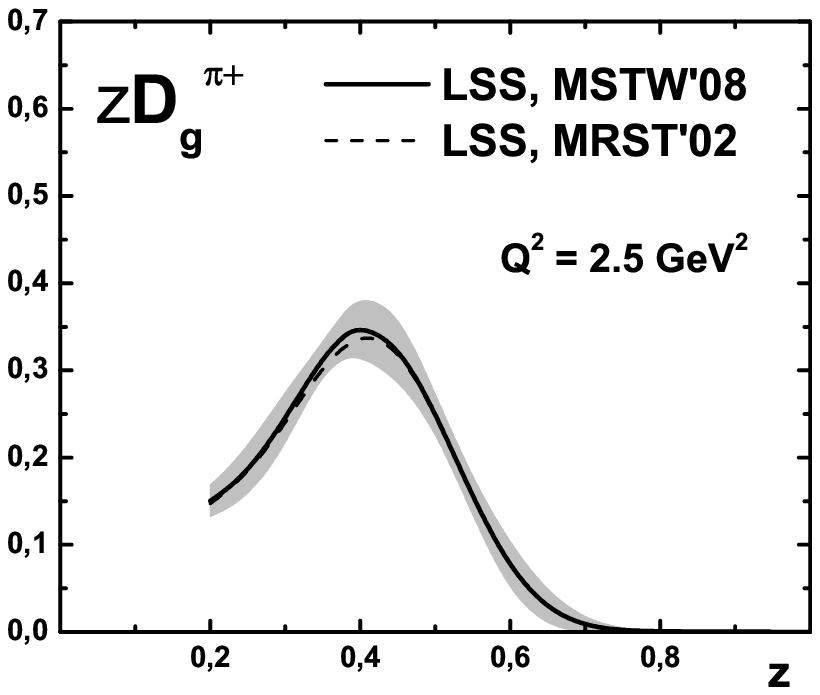}
\caption{\footnotesize Sensitivity of the extracted favored and unfavored FFs (left), and gluon FF (right) to the choice of the set of
unpolarized PDFs (see the text). Note the extremely small scale of
the vertical axis in Fig.~4(left). \label{fig4} }
\end{center}
\end{figure}

We have tried to get a feeling for the dependence of the results
on the unpolarized PDFs used in the analysis, and find that when
the MRST'02 set is used instead of the NLO MSTW'08 the description
of the data is slightly worse, with a value of $\chi^2/{\rm
d.o.f}$ equal to 1.00 (0.94 for MSTW'08 PDFs). In Fig.~\ref{fig4}
we illustrate the sensitivity of the extracted pion FFs to the use
of different sets of NLO unpolarized PDFs, in our case MWST'08 and
MRST'02. The corresponding FFs $D^{\pi^{+}}_{u}(z)$ and
$D^{\pi^{+}}_{\bar u}(z)$ are not shown because the differences
between them are so small that they are not visible. Instead, for
them, the error bands corresponding to the uncertainty estimates
at 68\% C.L. for $D^{\pi^{+}}_{u}(\rm MRST'08)$ (the black solid
curves) and $D^{\pi^{+}}_{\bar u}(z)$ (the dashed curves),
respectively, are plotted in Fig. 4(left), and compared with the
differences $\Delta D^{\pi^{+}}_{u}$ and $\Delta D^{\pi^{+}}_{\bar
u}$, short dash and dash dot dot curves, respectively, where
\begin{equation}
\Delta D^{\pi^{+}}_{u,\bar u} = D^{\pi^{+}}_{u,\bar u}(\rm
MRST'02) - D^{\pi^{+}}_{u,\bar u}(\rm MSTW'08). \label{diffFFs}
\end{equation}

Also for the gluons the difference is very small, but is at least
visible, as shown in Fig. 4(right). Note that because of the large
uncertainty in the determination of the gluon FF in Fig. 4(right)
only the error band corresponding to $\Delta \chi^2=1$ is
presented. As seen from Fig.~\ref{fig4}, the central values of the
fragmentation functions corresponding to the use of MRST'02 PDFs
lie entirely within the error bands for FFs corresponding to the
use of MSTW'08 set of PDFs. The fact that a choice of PDFs other
than the MSTW'08 set does not substantially alter the results of
the global fit was mentioned also in \cite{newDSS_FFs}. Thus, to
summarize, the extraction of the FFs is weakly dependent on the
choice of unpolarized PDFs.

\begin{figure} [t]
\begin{center}
  \includegraphics[height=.69\textheight]{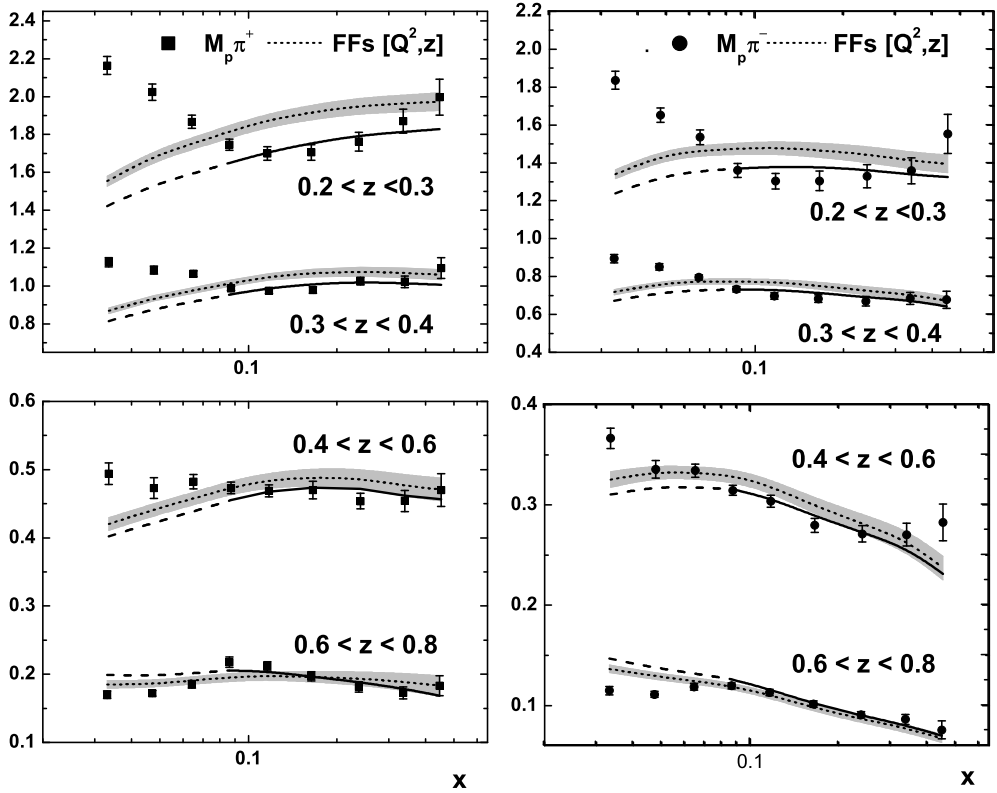}
\vspace{-5.0cm}
\caption{\footnotesize Comparison of HERMES $[x,z]$ {\it proton} data on
$\pi^{+}$ (left) and $\pi^{-}$ multiplicities (right) with the
multiplicities at the same kinematic points calculated by our FFs
extracted from HERMES $[Q^2,z]$ data (dot curves with the bands
corresponding to uncertainty estimates at 68\% C.L). The errors of
the data are {\it total}, statistical and systematic taken in
quadrature. The solid curves correspond to the best {\it fit} to
the cut $[x,z]$ data (see the text).\label{fig5} }
\end{center}
\end{figure}

\begin{figure} [t]
\begin{center}
\includegraphics[height=.69\textheight]{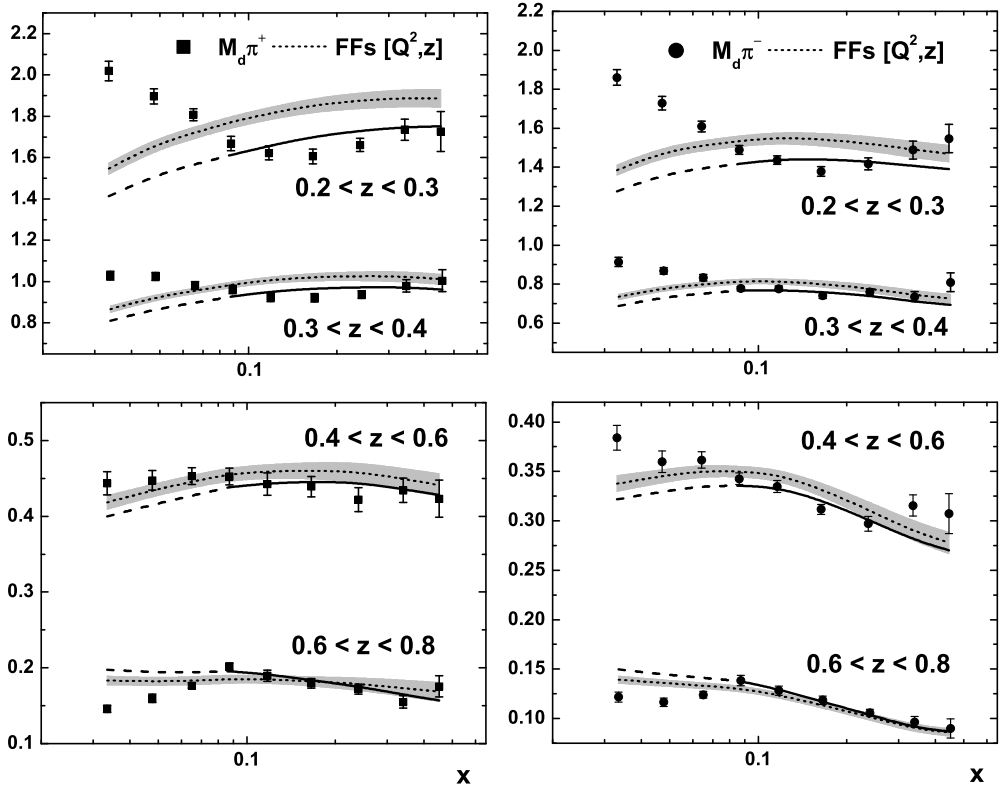}
\vspace{-5.0cm}
\caption{\footnotesize As in Fig. 5 but for a deuteron target.
\label{fig6}}
\end{center}
\end{figure}

Using the extracted FFs from the HERMES data on multiplicities in
the $[Q^2,z]$ presentation we have calculated the multiplicities
at the kinematic points for the data in the $[x,z]$ presentation.
The obtained value for $\chi^2$ is huge, 2187.8 for 144
experimental points (recall that the corresponding value of
$\chi^2$ for the $[Q^2,z]$ data is 123.95). The results are shown
in Fig.~\ref{fig5}  for the proton and in Fig.~\ref{fig6} for the
deuteron target. The theoretical multiplicities are presented
along with their uncertainty estimates corresponding to 68\% C.L.
As seen from the figures, the discrepancy is very large for both
the proton and deuteron targets for the first two z-bins [0.2-0.3]
and [0.3-0.4], as well as at lowest x, for all z-bins. In our
opinion such a significant discrepancy is totally unphysical. In
an attempt to understand this we have tried to fit the HERMES
$[x,z]$ data \emph{directly} and found that we cannot obtain a fit
with a reasonable $\chi^2$ using different input parametrizations
for the fragmentation functions. In addition, for some of the
parameters we obtain values in the non-physical region. It is
clear that the trend of the data in the small $x$ region is
different not only from that of the QCD predictions in this
region, but also from the rest of the data points in each z bin.
Consequently we decided to perform a NLO QCD fit to $[x,z]$ data
after removing the three lowest $x$ data points for every z bin.
 The total number of removed data points for
$\pi^{+}$ and $\pi^{-}$ multiplicities is 48 for which the
contribution to $\chi^2$ above is 1470 (30.6 per point). In the
fit to the rest of the data (96 points; we will refer to this data
set as the "cut" $[x,z]$ data) we have used for the input FFs the
parametrization given in Eq. (\ref{inputFFs}). Not unexpectedly it
turned out that the input parameters for the gluon FF can not be
fixed well from the fit, so for them we have used the parameters
obtained from the fit to the $[Q^2,z]$ data (see Table I). The
following value for $\chi^2/{\rm d.o.f}$, $\chi^2/{\rm d.o.f}$ =
179.37/87 = 2.06 for 96 experimental points and 9 free parameters,
is achieved in the fit. The results of the best fit are shown in
Fig. 5 for a proton target and Fig. 6 for the deuteron one (solid
curves). Their continuation to the low x region where the data
points were removed from the fit, is indicated by the dashed
curves. The quality of this fit is illustrated in Table II, and
compared to the quality achieved in the fit to the $[Q^2,z]$ data.
It follows from the $\chi^2$ values, presented in Table II, that
the description of the $[Q^2,z]$ data is much better than that of
the $[x,z]$ data even after removing a third of the data points.

The extracted pion favored and unfavored FFs from the fit to the
HERMES cut $[x,z]$ data on pion multiplicities are presented in
Fig. 7 and compared to those determined from the fit to $[Q^2,z]$
data, for which the error bands corresponding to the uncertainty
estimates at 68\% C.L. are also presented. Recall that the gluon
FF is the same for both the representations of the data and it is
shown in Fig. 3. As seen from Fig. 7, the central values of the
favored pion FF (solid curve) extracted from the cut $[x,z]$ data
are systematically smaller than those extracted from $[Q^2,z]$
data, and in the z region [0.2, 0.4] the corresponding curve lies
outside the error band. The central values of unfavored pion FF
extracted from the cut $[x,z]$ data are also systematically
smaller then those extracted from $[Q^2,z]$ data, however, the
corresponding curve lies within the error band. It is important to
mention, however, that from the calculation of the multiplicities
at the kinematic points for the data in the $[Q^2,z]$
presentation, using the extracted FFs from the fit to the cut
$[x,z]$ data, we obtain for $\chi^2$ the value 665.2 which is more
than five times larger than the value 123.95 achieved in the
direct fit to the $[Q^2,z]$ data.
\vskip 0.5cm
\begin{center}
\begin{tabular}{cl}
&{\bf TABLE II.}  $\chi^2$ per point values for the pion
multiplicities \\& obtained from the fits to $[Q^2,z]$
and the cut $[x,z]$ data.
\end{tabular}
\vskip 0.3 cm
\begin{tabular}{|c|c|c|c|} \hline
 &~~~$[Q^2,z]$ fit~~~&$~~~[x,z]$ fit ~~~ \\\hline
 $~~~M_p^{\pi^{+}}~~~$ &~~~ 0.83~~~ &~~~ 1.94 ~~~ \\
 $M_p^{\pi^{-}}$ & 0.65 & 1.58 \\
 $M_d^{\pi^{+}}$ & 0.98 & 1.63\\
 $M_d^{\pi^{-}}$ & 0.98 & 2.33\\ \hline
\end{tabular}
\end{center}
\vskip 0.5cm

Note that in all our NLO QCD calculations of the $[x,z]$ pion
multiplicities we have used for $x$ and $Q^2$ their mean values
$<x>$ and $<Q^2>$ as given in the HERMES data tables
\cite{HERMES}. We have checked that in NLO QCD the pion
multiplicities calculated at the average kinematics $\{<x>,
<Q^2>\}$ coincide extremely closely (to better than 1\%) with the
average multiplicities calculated using the expression, Eq. (1),
given in the recent HERMES paper \cite{HERMES_Reply} as applied to
the NLO semi-inclusive and DIS cross sections (see the remark
\cite{comment}). This fact is very important because it means that
the huge time consuming the computer calculations involved in
using the above mentioned expression in fitting the data on the
average multiplicities can be significantly reduced if the NLO QCD
multiplicities are calculated at the corresponding mean values
$<x>$ and $<Q^2>$.

Finally we would like to underline that our NLO QCD analysis of
the HERMES $[x,z]$ data supports the assertion of Stolarski
\cite{Stolarski}, based on a LO QCD analysis, that the increase of
magnitude of the HERMES pion multiplicitity sum as $x$ decreases
in the region $x < 0.1$, is difficult to reconcile with
perturbative QCD. While in \cite{Stolarski} the argument is
presented for a deuteron target, our observation is that it holds
for both proton and deuteron targets.

\begin{figure}[bht]
\begin{center}
\includegraphics[scale=0.70]{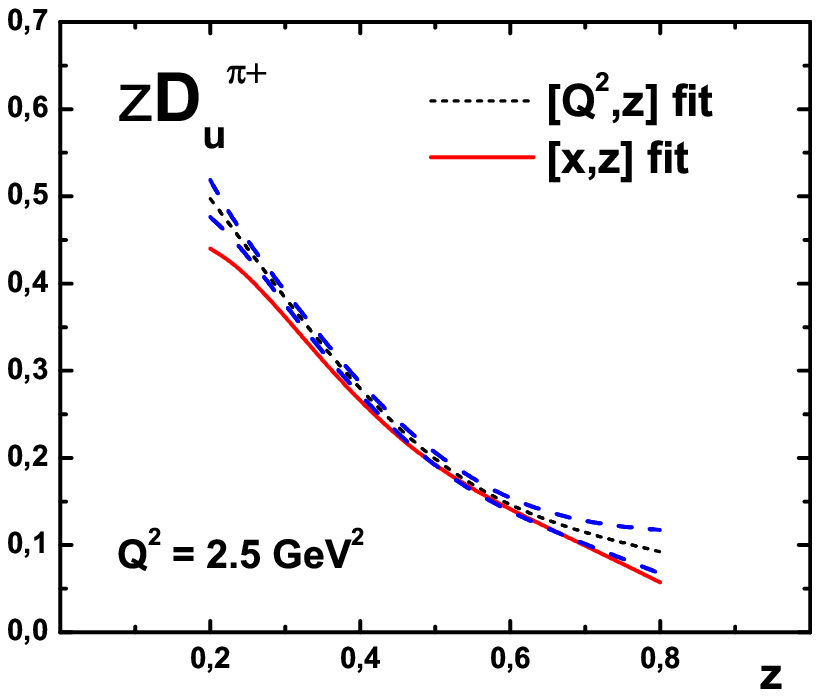}
\includegraphics[scale=0.70]{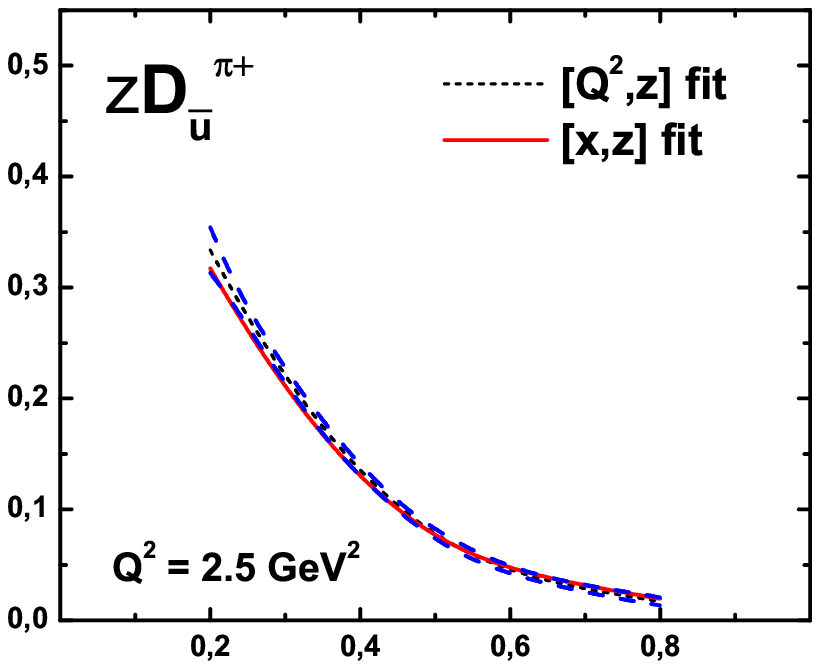}
\caption{Comparison between favored (left) and unfavored (right)
fragmentation functions extracted from $[Q^2,z]$ (dot curves) and
the cut $[x,z]$ HERMES data on multiplicities (solid curves). The
dashed curves mark the error bands for FFs determined from the
$[Q^2,z]$ data. \label{FFs_q2z_vs_xz} }
\end{center}
\end{figure}

\section{Summary}
The publication by HERMES of the final version of their data on pion
multiplicities on protons and deuterons has profound implications for our
understanding of the pion fragmentation functions.

1) The fact that the final data are significantly different from the preliminary
data means that the oft utilized DSS FFs \cite{DSS}, which were based on the
preliminary data, are incorrect.

2) We have studied the two-dimensional projections of the final
HERMES data, the so-called $[x,z]$ and $[Q^2,z]$ formats,
presented by the HERMES group.\\
$~~~~~$a) With the pion FFs, parametrized in a standard way, and
respecting isospin invariance, we have found an excellent fit to
the $[Q^2,z]$ presentation of the data and extracted a new set of
NLO pion FFs. Except for the gluon fragmentation function, our new
pion FFs are very similar indeed to those obtained recently by the
DSEHS group \cite{newDSS_FFs} using, in their global analysis,
the $[Q^2,z]$ HERMES data.\\
$~~~~~$b) On the contrary, no reasonable NLO QCD fit could be
achieved to the $[x,z]$ presentation of the data. We have found
that an adequate fit to the $[x,z]$ data is only possible if we
cut points with $x < 0.075$ from the data which means that a third
of the data points is removed.  However, even with these cuts, the
quality of the description of the $[Q^2,z]$ data is much better
than that achieved for the cut $[x,z]$ data. While the extracted
unfavored pion FF lies within the error band corresponding to the
unfavored pion FF extracted from the fit to the $[Q^2,z]$ data,
the favored pion FF$[x,z]$ is systematically smaller than favored
FF$[Q^2,z]$ and is outside of its error band in the region $0.2 <
z < 0.4$.

3) We have found that the trend of the data in the HERMES $[x,z]$
presentation of their data, where the magnitude of the pion
multiplicitities in the region $x < 0.1$ increases as $x$
decreases, is totally at variance with the trend of the  NLO QCD
predictions. This suggests that possibly there is a problem with
the HERMES $[x,z]$ presentation of their  data, and emphasizes the
need for new data on the hadron multiplicities. We thus await with
great interest the publication of the final COMPASS data on the
pion multiplicities.

\begin{center}
{\bf Acknowledgments}
\end{center}

We are grateful to M. Stolarski for the useful discussions.
This research was supported by the JINR-Bulgaria Collaborative Grant,
and by the Russian Foundation for Basic Research Grants No.
13-02-01005 and No. 14-01-00647. E. L. is grateful to the Leverhulme
Trust for an Emeritus Fellowship.


\begin{thebibliography}{99}
\bibitem{comment}
Note that Eq. (1) for the multiplicities in the $[x,z]$
presentation of the HERMES data as given in \cite{HERMES_Reply},
is not correct. It disagrees with the verbal prescription given in
\cite{HERMES_Reply} as to how to use the HERMES database on the
multiplicities, in order to compare the experimantal value of the
$[x,z]$ pion multiplicity measured in a given $x$-bin to the
theory predictions for the same bin. According to this
prescription one has to integrate the numerator and denominator in
Eq. (1) over all $x$ values lying between the lower and upper
boundaries of the bin.
\end{thebibliography}
\end{document}